\documentclass[
    a4paper,
    11pt
]{article}
\pdfoutput=1 

\usepackage{jinstpub} 

\usepackage{lineno}

\usepackage{siunitx}

\usepackage{caption}
\usepackage{subcaption}

\usepackage{listings}

\usepackage{mathtools}

\usepackage{multirow}

\usepackage{dirtytalk}

\usepackage{hyperref}

\usepackage{tabularx}

\title{SiPM Signal Processing via Multiple Linear Regression }

\author[a,b,1]{Wolfgang Schmailzl,\note{Corresponding author.}}
\author[b]{Claudio Piemonte,}
\author[c]{Erika Garutti}
\author[a]{and Walter Hansch}

\affiliation[a]{Institute of Physics, Bundeswehr University Munich, Germany}
\affiliation[b]{Broadcom Inc., Regensburg, Germany}
\affiliation[c]{University of Hamburg, Hamburg, Germany}

\emailAdd{wolfgang.schmailzl@gmail.com}

\abstract{
This paper presents a novel approach using multiple linear regression to process transient signals from silicon photomultipliers.
The method provides excellent noise suppression and pulse detection in scenarios with a high pulse count rate and superimposed pulses.
Insights into its implementation and benchmark results are presented.
We also show how this approach can be used to automatically detect the pulse shape from a given transient signal, providing good detection for count rates up to \SI{90}{MHz}.
Experimental data are used to present an application where this algorithm improves charge spectrum resolution by an order of magnitude.
}


\begin{document}

\renewcommand{\labelenumii}{\arabic{enumi}.\arabic{enumii}}
\renewcommand{\labelenumiii}{\arabic{enumi}.\arabic{enumii}.\arabic{enumiii}}
\renewcommand{\labelenumiv}{\arabic{enumi}.\arabic{enumii}.\arabic{enumiii}.\arabic{enumiv}}

\maketitle
\flushbottom

\section{Introduction}

Silicon photomultipliers (SiPMs) are an excellent choice for many photodetection applications \cite{gundacker_silicon_2020}.
Such applications include positron emission tomography \cite{lecoq_sipm_2021}, bio-sensing \cite{santangelo_si_2016}, time tagging of high energetic particles \cite{garutti_silicon_2011} or light detection and ranging \cite{bilik_comparative_2022,acerbi_high-density_2018}.
Signal shape, dark count rate, afterpulsing, and crosstalk can vary widely between SiPMs and operating conditions..
Therefore, proper characterization and data evaluation methods are required to select the right SiPM for different applications or to improve and better understand its behavior in different environments.
The choice of readout electronics also affects the response of the SiPM \cite{acerbi_understanding_2019}. This includes electronic bandwidth, noise level or signal amplitude. 

Efficient pulse detection is the first and arguably most important step in processing SiPM signals.
The easiest way is to set a threshold on the raw signal and look for pulses. 
This procedure is inaccurate if a subsequent pulse occurs during the recovery time of the preceding pulse.
Such superimposed pulses become more likely as the pulse count rate increases.
Signal filtering can greatly improve pulse identification, and several approaches have been considered in recent years.
A common filter type for pulse detection is the trapezoidal filter. It provides good noise reduction and an efficient implementation \cite{stein_x-ray_1996,nakhostin_signal_2018}.
The trapezoidal filter type used for SiPM pulse processing has been demonstrated in \cite{engelmann_sipm_2018}.
However, filter lengths and weights must be tuned for each pulse shape, and dead time can affect accuracy at high pulse count rates.
In \cite{piemonte_development_2012,gola_dled_2012}, a differential leading edge discriminator method with subsequent peak detection is used to evaluate SiPM signals.
Another approach is based on a piecewise linear fit \cite{putignano_non-linear_2012}, where the studied curve is approximated with straight lines of different lengths and gradients.
In \cite{bychkova_radiation_2022}, a moving window average and moving windows differential are combined with a moving windows searching algorithm for SiPM pulse detection. 

This paper presents a novel approach using multiple linear regression to process transient SiPM signals.
We present two algorithms, the first is the core, which iteratively detects pulses in the transient signal, their arrival times, and the amplitudes of each pulse, even in the case of highly superimposed pulses.
This algorithm requires a pulse shape as user input.
We address this limitation with the second algorithm, which can automatically determine the pulse shape from the transient signal.
In terms of hardware requirements, our algorithms can run on any system capable of digital signal processing and solving linear systems of equations.

Section~\ref{sec:description} first defines the mathematical context for using multiple linear regression for SiPM signal processing. Then, the additional subroutines of the algorithm are introduced and the overall processing procedure is described.
The last part of this section introduces the pulse shape recognition algorithm.
In section~\ref{sec:performance}, we first look at implementation options to improve the runtime and present benchmarks.
Then, the performance of the algorithm is examined using simulated and experimentally measured data.
Charge spectra of the processed experimental data are presented to show an application example where our algorithm improves the resolution compared to a standard method.

\section{Algorithm Description}
\label{sec:description}

We use multiple linear regression to reconstruct a transient SiPM signal containing at least one SiPM pulse, i.e. Geiger discharge.
Our algorithm can be used in two ways: 
\begin{enumerate}
    \item The algorithm requires a pulse shape as an input, returns the pulse positions and solves for their amplitudes.
    \item The pulse shape is unknown and the algorithm determines it automatically.
\end{enumerate}
The idea of this approach is based on the concept that the residual between data and model is minimized if the pulse shape is placed and superimposed in the reconstructed signal at all arrival times of the actual pulses. 
The amplitudes of each pulse are then fitted by least squares.

\begin{figure}
    \centering
    \includegraphics[width=0.9\columnwidth]{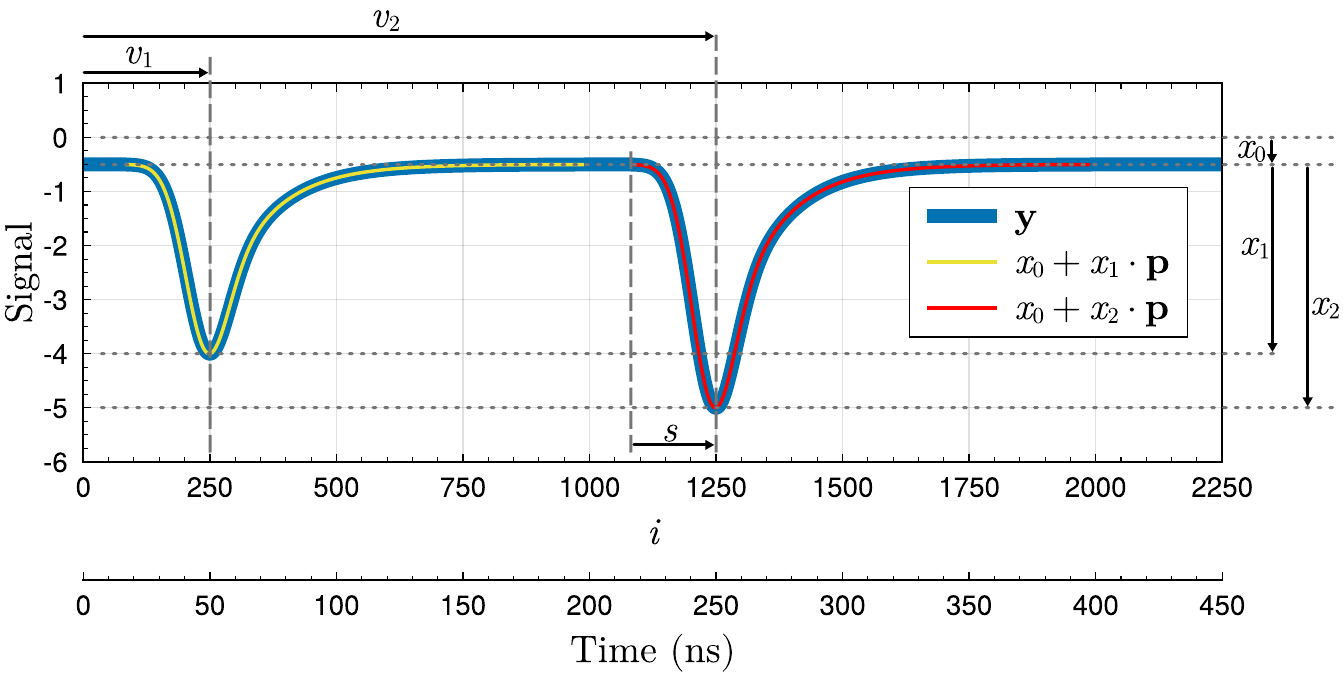}
    \caption{Example showing how to construct a signal vector $\mathbf{y}$ from two pulse vectors $\mathbf{p}$, plotted as a function of its entry index $i$ (or as a function of time for a sampling rate of \SI{5}{GS/s}). The pulses are shifted by $v_1=250$ and $v_2=1250$ in the horizontal direction with respect to the peak position ($s=170$) and by $x_0=-0.5$ in the vertical direction. The pulse scaling is done using $x_1=-3.5$ and $x_2=-4.5$. The remaining entries of $\mathbf{y}$ are filled with $x_0$ due to the limited length of $\mathbf{p}$.}
    \label{fig:math_explain}
\end{figure}

Our algorithm uses simple peak detection to first estimate the positions of the pulses. This initial estimate is used to solve for the amplitudes to reconstruct the signal by linear regression.
Pulse positions are improved iteratively and initially undetected pulses can be detected by remaining peaks in the residual.
An overview of the required variables and an example of their contextual meaning is given in table~\ref{table:definitions}, with detailed descriptions in subsection~\ref{sec:math}.
Required user input is the signal vector $\mathbf{y}$ and the pulse vector $\mathbf{p}$, while all other variables are determined by the algorithm.
In figure~\ref{fig:math_explain}, a transient signal with two pulses is presented and the effect of each variable is visualized.

The limitation of the required pulse shape input is addressed in subsection~\ref{sec:shape_recognition}, where we present a method to automatically recognize the pulse shape from the transient input signal.

\begin{table}
    \centering
    \caption{Overview of all required variables, their description and examples of contextual meaning.}
    \begin{tabularx}{\textwidth}{l l X }
        \hline
        Variable        & Description       & Contextual Meaning \\ [1ex]
        \hline
        $\mathbf{y}$    & Signal vector     & Single waveform acquired with an oscilloscope at a constant sampling rate, e.g. the time-dependent voltage at the SiPM preamplifier output.   \\ [0.8ex]
        $\mathbf{p}$    & Pulse vector      & Pulse shape data (same sampling rate as $\mathbf{y}$) containing only non-zero values. Represents an averaged SiPM single-photon response and is typically normalized to one. \\ [0.8ex]
        $\mathbf{v}$    & Pulse position vector & Each entry of this vector represents the position of a pulse in $\mathbf{y}$, where the number of entries is equal to the total number of pulses. \\ [0.8ex]
        $s$             & Pulse peak position & Position of the peak in $\mathbf{p}$, where $1 \leq s \leq \text{dim}(\mathbf{p})$. \\ [0.8ex]
        $\mathbf{x}$    & Pulse amplitude vector & The number of entries of this vector is $\text{dim}(\mathbf{v})+1$. The first entry represents the offset voltage $x_0$ of the waveform, and if $\mathrm{max}(\mathbf{|p|})=1$, then each subsequent entry represents the amplitude of a pulse already subtracted from $x_0$.\\ [0.8ex]
        \hline
    \end{tabularx}
    \label{table:definitions}
\end{table}

\subsection{Mathematical description}
\label{sec:math}

Let us define a vector $\mathbf{y}$ representing an equidistantly sampled transient signal generated by a SiPM:
\begin{equation}
    \mathbf{y} = \begin{pmatrix}y_1 & y_2 & \dots & y_{I-1} & y_{I}\end{pmatrix}^\top,
\end{equation}
provided that $\mathbf{y} \in \mathbb{R}^I$ and where $I$ is the number of samples per signal.
As an example, this can be the vector of samples acquired by an oscilloscope containing several SiPM pulses in dark conditions.
As an initial assumption, the exact response of the SiPM to a Geiger discharge is known and this pulse shape is stored in the vector
\begin{equation}
    \mathbf{p} = \begin{pmatrix}p_1 & p_2 & \dots & p_{K-1} & p_{K}\end{pmatrix}^\top,
\end{equation}
given that $\mathbf{p} \in \mathbb{R}^K$ and $p_{1,K} \neq 0$.
Additionally, a pulse should provide a unique peak position at
\begin{equation}
    s = \underset{k}{\mathrm{arg\,max}}  \, |p[k]|,
\end{equation}
where $s \in [1, K]$.
Typically, a transient signal of a SiPM consists of a certain number $J$ of individual pulses. The information about the position of each pulse in the signal vector is stored in
\begin{equation}
    \mathbf{v} = \begin{pmatrix}v_1 & v_2 & \dots & v_{J-1} & v_{J}\end{pmatrix}^\top,
\end{equation}
where $\mathbf{v} \in \mathbb{N}^J$. Furthermore, each pulse has a variable amplitude with
\begin{equation}
    \mathbf{x} = \begin{pmatrix}x_1 & x_2 & \dots & x_{J-1} & x_{J}\end{pmatrix}^\top
\end{equation}
and $\mathbf{x} \in \mathbb{R}^J$.

We can now define a simplified model function consisting of only a single pulse to generate a reconstructed signal vector. 
The peak of the single pulse is placed at the position $v$ using $s$ and with an amplitude of $x$,
\begin{equation}
    g(n; x, v) = 
    \begin{cases}
        x \cdot p[n-v+s],     & \text{if } (-s+1+v) \leq n \leq (K-s+v)\\
        0,                      & \text{otherwise.}
    \end{cases}
\end{equation}
This single pulse model function can be extended to a multiple pulse model function using the previous definitions, leading to
\begin{equation}
    f(n; \mathbf{v, x}, x_0) = x_0 + \sum_{j=1}^J g(n;x[j], v[j]),
\end{equation}
where $x_0$ is an offset. In the context of transient SiPM signals, this can be a voltage offset due to a pre-amplifier. Another way to describe the model function is in matrix-vector representation and, if $\mathbf{x}$ is extended by $x_0$, we get
\begin{equation}
    \label{eq:Atimesx}
    \mathbf{A \cdot x} = \begin{pmatrix}
        1 & g(1;1, v[1]) & g(1;1, v[2]) & \cdots & g(1;1, v[J]) \\
        1 & g(2;1, v[1]) & g(2;1, v[2]) & \cdots & g(2;1, v[J]) \\
        \vdots & \vdots  & \vdots       & \ddots & \vdots       \\
        1 & g(I;1, v[1]) & g(I;1, v[2]) & \cdots & g(I;1, v[J]) \\
    \end{pmatrix} 
        \cdot 
    \begin{pmatrix}
            x_0 \\
            x_1 \\
            x_2 \\
            \vdots \\
            x_J
    \end{pmatrix}.
\end{equation}
In other words, the rectangular matrix $\mathbf{A}$ of the size $I \times (J+1)$ contains in each column, expect the first one, $\mathbf{p}$ shifted to the respective position defined by $\mathbf{v}$. 
The vector $\mathbf{x}$ represents the offset and all pulse amplitudes. 
In practice, the pulse length is typically much shorter than the signal length, so most entries of $\mathbf{A}$ are zero resulting in a sparse matrix.

\begin{figure}
    \centering
    \includegraphics[width=0.9\columnwidth]{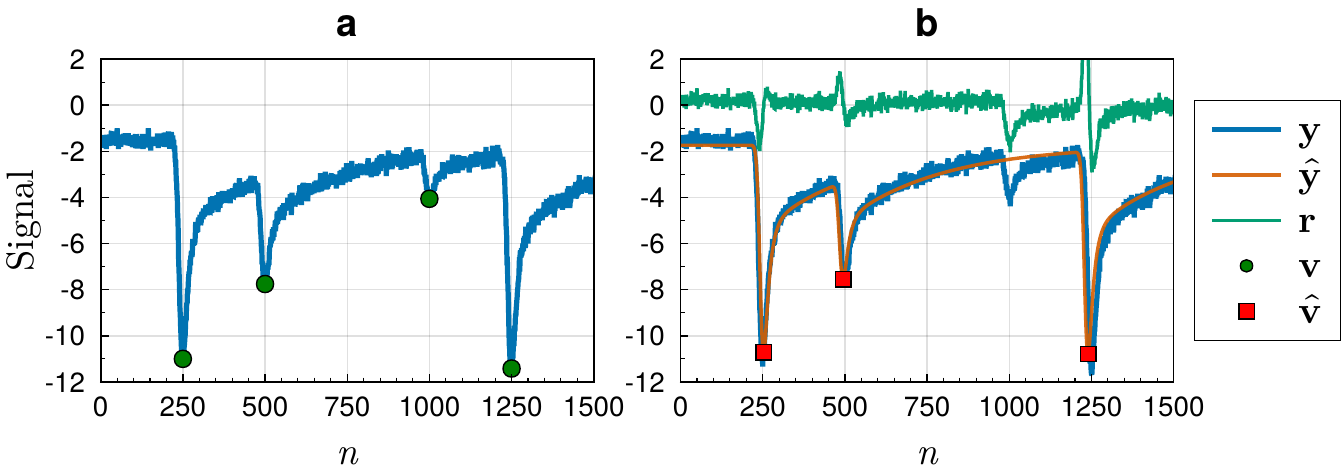}
    \caption{\textbf{(a)} Example of a signal with a SiPM-like pulse shape and superimposed white noise, generated using $\mathbf{A \cdot x}$. \textbf{(b)}~Least squares solution with slightly misaligned or missing pulse positions.}
    \label{fig:model_function}
\end{figure}

In a scenario where the pulse positions are known or already estimated, we can construct $\mathbf{A}$ while $\mathbf{x}$ is the only unknown.
However, $\mathbf{A \cdot x = y}$ has no solution ($I$ greater than $J$), but we can compute
\begin{equation}
    \label{eq:minAx=y}
    \text{min} \; ||\mathbf{A \cdot x - y}||^2
\end{equation}
to find a least squares solution $\mathbf{\hat{x}}$ for the amplitudes and to get the reconstructed signal $\mathbf{\hat{y}= A \cdot \hat{x}}$. The residual vector is then $\mathbf{r} = \mathbf{y} - \mathbf{\hat{y}}$.

To prove our methodology, we can define our own $\mathbf{A}$ and $\mathbf{x}$ to generate simulated SiPM signals.
Noise can be added if necessary.
As an example, figure~\ref{fig:model_function}a shows a signal generated with this approach. 
In this case the position vector is $\mathbf{v}=(250, 500, 1000, 1250)^\top$ and the offset/amplitude vector is $\mathbf{x}=(-1.5, 9.5, 4.6, 1.9, 9.3)^\top$. 
White noise with a sigma of 0.2 is added.

In figure~\ref{fig:model_function}b, an example of a least-square solution is shown where estimated pulse positions ($\mathbf{\hat{v}}$) are used. 
This example presents two problems that may occur: Undetected and misplaced pulse positions.
The first issue can be seen at the third smaller peak, which is not detected. The second problem is especially noticeable with the fourth peak. Here the position estimation is slightly wrong, resulting in a smaller reconstructed amplitude. 
In the following two sections, we will first look at how to optimize each pulse position, and in the next section, how to deal with undetected or erroneously detected pulses.

\subsection{Pulse position optimizer}

\begin{figure}[]
    \centering
    \includegraphics[width=0.9\columnwidth]{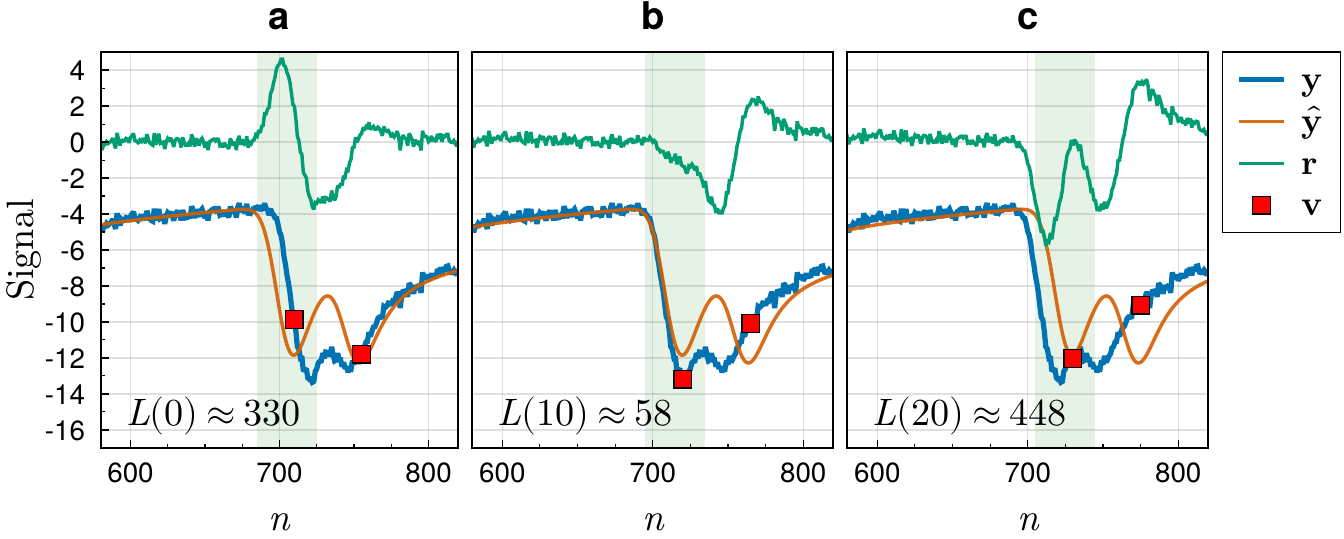}
    \caption{ A single pulse can be shifted efficiently without recomputing $\mathbf{\hat{y}}$. The residual vector in a specified optimization interval (light green) around the current pulse position is used as input to the loss function to find the optimized position. \textbf{(a)}~Initial positions and pulse heights solution solving equation~\ref{eq:minAx=y} to get $\mathbf{\hat{y}}$. \textbf{(b)}~First pulse shifting without height recomputation (new optimum). \textbf{(c)}~Second pulse shifting without height recomputation.}
    \label{fig:pulse_position_optimization}
\end{figure}

As mentioned above, linear least-squares fitting is used to evaluate the individual pulse amplitudes.
However, positioning the pulses is challenging and, if not done correctly, can cause errors in the fitted amplitudes.
This amplitude error can even propagate to subsequent pulses if pulses are superimposed as shown in figure~\ref{fig:model_function}b.

We start the optimization process with the initial guess of the pulse positions $\mathbf{v}$ and the least-square solution $\mathbf{\hat{y}}$.
Each pulse position is processed individually, and amplitudes are not recalculated to reduce computational overhead.
We define an interval relative to the pulse position of interest to select a subset of $\mathbf{\hat{y}}$ that we will use for the optimization. 
For example, if the pulse position to be processed is $v_3$, and 25 entries to the left and 15 entries to the right of that position are selected, then the subset would be $\left( \hat{y}_{v_3-25} \, \dots \, \hat{y}_{v_3+15} \right)$.
This subset is then shifted over $\mathbf{y}$, and the new optimized position is found where a shift-dependent loss function has its minimum.
Let $n_s$ be the shift variable here, and we choose a squared error loss function, then the new position in this example would be where
\begin{equation}
    L(n_s) =\sum_{i=v_3-25}^{v_3+15} (y[i+n_s] - \hat{y}[i])^2
\end{equation}
is minimal.
This process is then repeated for all other pulses in the signal. 

A visualized example of this pulse position optimization with two shift operations is shown in figure~\ref{fig:pulse_position_optimization} and we optimize the pulse position on the left.
The initial guess is shown in figure~\ref{fig:pulse_position_optimization}a. 
It is visually clear that this first estimation of both pulse positions is not satisfactory.
The loss of this pulse is evaluated in the interval visualized by the vertical green band.
Then a first shift is applied and we can see how the residual has changed in figure~\ref{fig:pulse_position_optimization}b, resulting in a significantly lower loss.
Shifting even further to the right now increases the loss as shown in figure~\ref{fig:pulse_position_optimization}c. 
So for the first pulse, the updated position will be the one shown in figure~\ref{fig:pulse_position_optimization}b. At this point, the second pulse is considered with the same procedure.

After all pulse positions have been processed, $\mathbf{\hat{y}}$ can be calculated again to update the amplitudes. 
The sequence of pulse position optimization and amplitude calculation can be repeated a fixed number of times or until the pulse positions no longer change.
Increasing the number of repetitions and the number of shifts per pulse position improves the results at the expense of run time.
Increasing the interval also increases the run time, and good results are often obtained when the interval is placed around rapid changes, such as the initial slope.

\subsection{Algorithm sequence}

\begin{figure}[]
    \centering
    \includegraphics[trim={15mm 423pt 15mm 5mm}, width=0.5\columnwidth,clip]{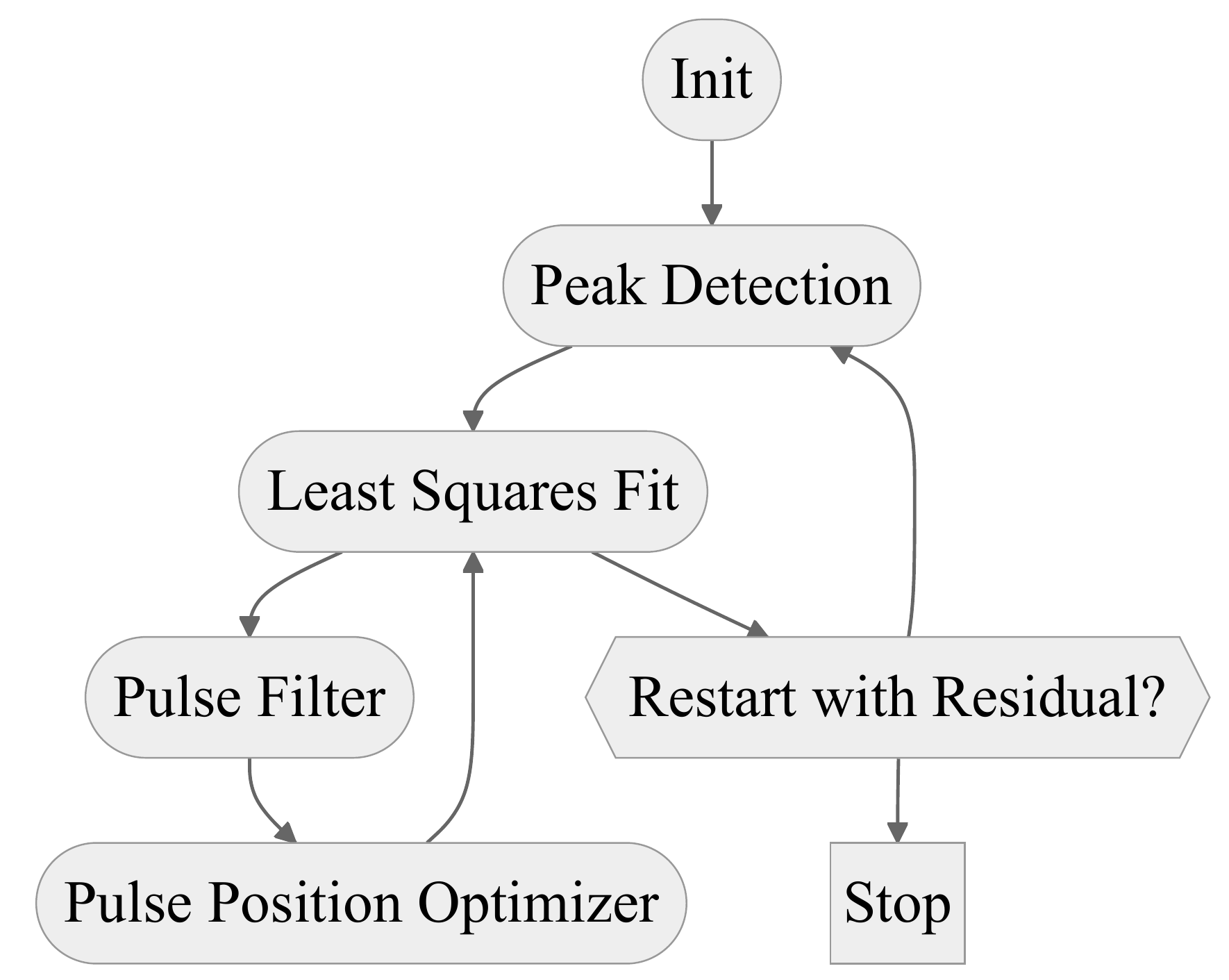}
    \caption{
        Simplified algorithm flow chart.
    }
    \label{fig:algorithm_flowchart}
\end{figure}

\begin{figure}[]
    \centering
    \includegraphics[width=0.9\columnwidth]{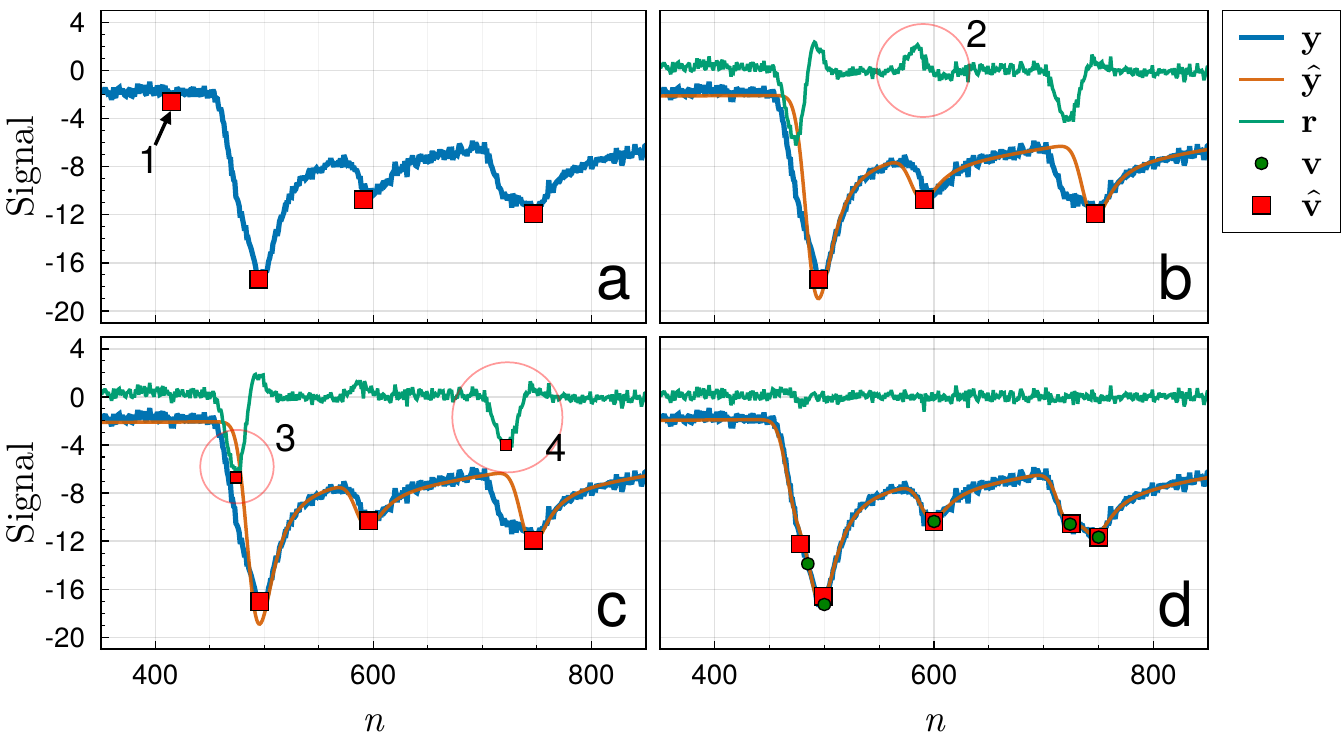}
    \caption{\textbf{(a)}~Inital peak detection - solver still inactive. \textbf{(b)}~First least squares solution. \textbf{(c)}~Pulse position optimization and peak detection on residual. \textbf{(d)}~Final result after two algorithm iterations.}
    \label{fig:algorithm_example}
\end{figure}

We now describe the complete algorithm and how we solve the problem of undetected or erroneously detected pulses.
A flowchart of the algorithm is shown in figure~\ref{fig:algorithm_flowchart}, and the effect of the different subroutines is visualized in the example signal vector presented in figure~\ref{fig:algorithm_example}.

In the \textbf{Init} state, the signal vector $\mathbf{y}$ and the pulse vector $\mathbf{p}$ including its peak position $s$ are passed to the algorithm.
The vectors $\mathbf{x}$ and $\mathbf{v}$ are empty at this stage.
Then following the \textbf{Peak Detection} subroutine, we use a method similar to \cite{brakel2014} to find peaks in the input vector passed to this function. In the first iteration of the algorithm the signal vector $\mathbf{y}$ is used as input, and all subsequent iterations use the residual vector $\mathbf{r}$.
After a single run of this function, there may be missed or erroneously detected pulses. 
The former can be detected in subsequent iterations and the latter can be filtered in the \textbf{Pulse Filter} subroutine.
Figure~\ref{fig:algorithm_example}a shows the state of the vectors after peak detection, where point \textbf{1} seems like a false positive due to noise.

In the \textbf{Least Squares Fit} subroutine, all variables of table~\ref{table:definitions} are now given, the matrix $\mathbf{A}$ is constructed and equation~\ref{eq:minAx=y} is solved.
The \textbf{Pulse Filter} can be used to remove pulses that do not meet a minimum amplitude.
This filtering approach is based on the assumption that the pattern of random noise has a low probability of matching the shape of the pulse, therefore resulting in a low least-squares amplitude.
The result of the first iteration after these steps is shown in figure~\ref{fig:algorithm_example}b, where the residual still shows three large perturbations. 

As described above, the \textbf{Pulse Position Optimizer} shifts each local pulse around its detected position to find a new optimum.
Figure~\ref{fig:algorithm_example}c shows the results of the updated pulse position vector and we can see that it has been able to reduce the residual of point \textbf{2} while points \textbf{3} and \textbf{4} remain almost unchanged.
A new iteration can be started now with the residual as input for the peak detection. 
Two new pulses have been detected at point \textbf{3} and \textbf{4} in figure~\ref{fig:algorithm_example}c. They are added to the previously detected pulses.
The internal loop of least squares fitting, pulse filtering and pulse position optimization can be entered again.

In this example, a satisfactory approximation of the true signal is obtained in this second iteration, as shown in figure~\ref{fig:algorithm_example}d.
A generated signal has been used here. Therefore, the true pulse positions $\mathbf{v}$ are known and are shown with green markers.
Despite the proximity of the two pulses at points 3 and 4, respectively, the fitted signal is a good approximation of the true signal.
However, a limitation of this approach can be seen in point 3.
Two pulses are so close together that noise affects the pulse position optimizer and the position error increases.
This error is a function of the pulse shape and the noise present.

\subsection{Pulse shape recognition}
\label{sec:shape_recognition}

The method of processing SiPM signals introduced in this paper was based on the assumption that the pulse shape is known, but this is rarely the case in practice.
In this section, we show how it can be automatically reconstructed using the previously described algorithm in an iterative process.

\begin{figure}[]
    \centering
    \includegraphics[width=1\columnwidth]{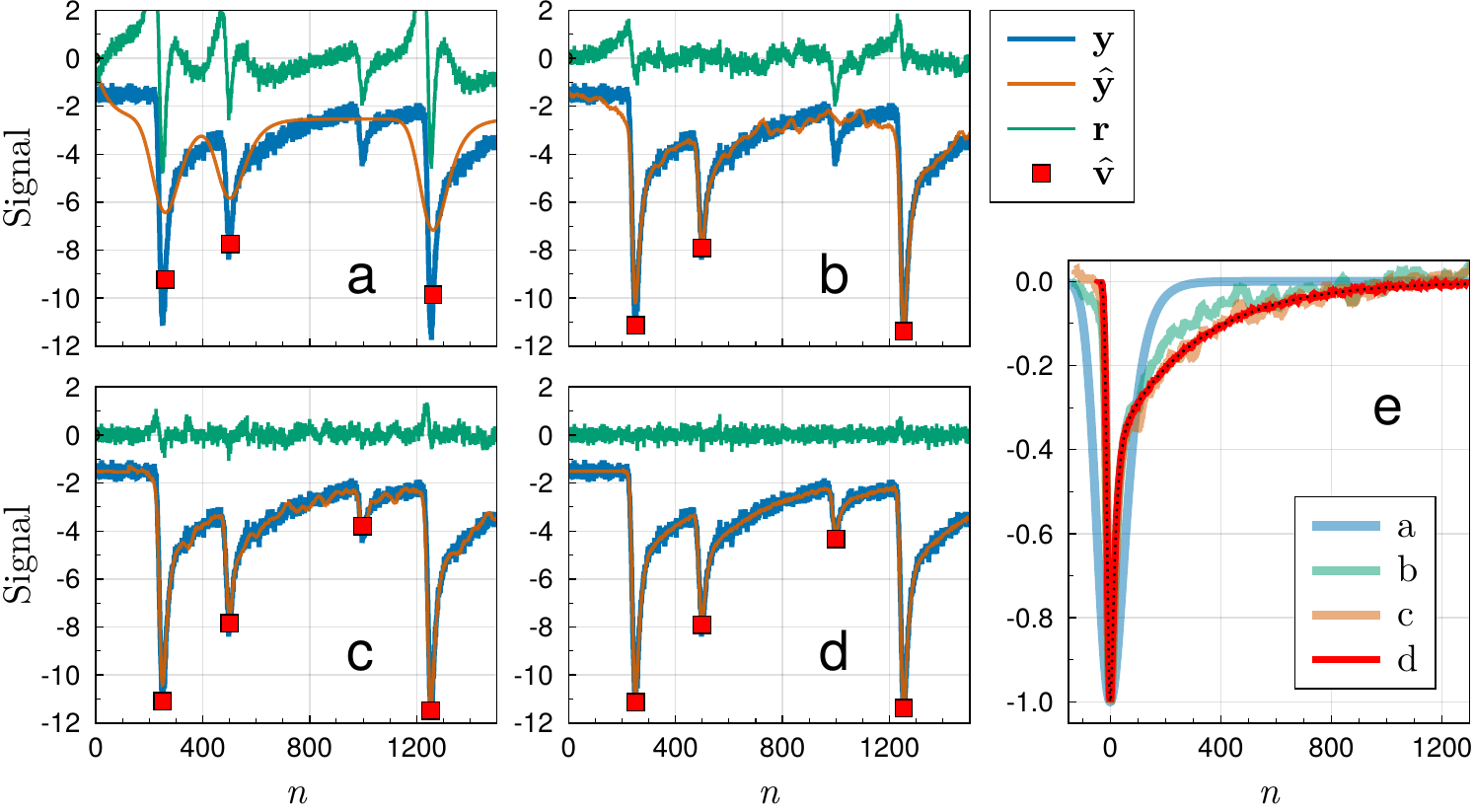}
    \caption{Example of the iterative pulse shape recognition. \textbf{(a)}~Inital guess using a Gaussian-shaped pulse. \textbf{(b)}, \textbf{(c)} and \textbf{(d)}: Fitted signal and estimated pulse shape after 5, 20 and 80 iterations, respectively. \textbf{(e)}~Pulse shape evolution with the true pulse shape in black. }
    \label{fig:pulse_shape_recognition}
\end{figure}

At startup, the actual pulse shape is still unknown. Therefore, an initial guess is required the first time the pulse processing algorithm (figure~\ref{fig:algorithm_flowchart}) is run.
Figure~\ref{fig:pulse_shape_recognition}a shows this first iteration using a Gaussian-like pulse shape as an initial guess.
Even if the shape of the pulse is not a good fit, the algorithm will still try to detect the pulse positions and minimize the residual.

Each detected pulse position is then used to update the shape.
The pulse vector is shifted along the signal vector so that at each iteration the peak of the pulse vector is aligned with the current pulse position: A peak-to-peak alignment. The section of the residual where the pulse vector is currently located is selected, multiplied by a learning rate, and then added to the pulse vector.
The learning rate (or update rate) dampens the effect of undetected or erroneous pulses that can cause high amplitudes in the residual before they are added to the current pulse shape.
However, a low learning rate requires more iterations.

After each pulse position is processed, the pulse shape is normalized so that $\mathrm{max}(\mathbf{|p|})=1$.
In addition, entries at the beginning or end of the pulse vector can be added or removed: 1)~Remove entries if there are too many leading or trailing zeros where no information is stored and only electronic noise accumulates in the pulse vector.
2)~Add entries if there are sharp transitions at the ends of the pulse shape vector.
So during runtime, the length of the pulse shape is variable.

Figure~\ref{fig:pulse_shape_recognition}b shows the change in pulse shape and the fitted signal after 5 iterations.
The first rising and falling edges compared to the true pulse shape are already well detected whereas the tail is still quite distorted.
As recognition progresses, the learning rate can be adjusted to ensure that the pulse shape converges.
In figure~\ref{fig:pulse_shape_recognition}c,d, the noise on the tail is further reduced, and after \SI{80}{iterations}, the detected pulse shape is very close to the true pulse shape in this example.
Figure~\ref{fig:pulse_shape_recognition}e shows how the pulse vector evolves as the number of iterations increases.
It should be noted that the quality of the initial guess will affect the number of iterations required and the convergence rate of the recognition. 
Also, if the initial guess leads to no pulse detections, this method will fail.

\section{Performance}
\label{sec:performance}

In this section, we discuss implementation options and present benchmark results for runtime and memory requirements.
The capabilities and limitations of the algorithms are presented using simulated and experimental data.

\subsection{Algorithm implementation and runtime}

The previous section explained how the algorithm works and demonstrated its capabilities.
An additional factor for the user of such an algorithm could be its runtime and how 
it scales with increasing signal length, pulse length or number of pulses per signal.
The most demanding operation is typically the least square fitting. 
Since $\mathbf{A \cdot x=y}$ has no solution, we instead solve the normal equation
\begin{equation}
    \label{eq:solve_inv}
    \mathbf{A^\top A \hat{x} = A^\top y}.
\end{equation}
To solve for $\mathbf{\hat{x}}$, several methods can be used \cite{strutz_data_2015,strang_introduction_2016}. For example, by inverting the normal matrix $\mathbf{A^\top A}$ or by using a matrix factorization.

To implement the algorithm described in the previous section, we use the programming language Julia \cite{bezanson_julia_2017} because of its high performance and features well suited to computational science.
Julia offers the backslash operator to solve linear problems and automatically determines which method to use \cite{juliadoc}, e.g. $\mathbf{\hat{x} = A \setminus y}$.
Additional runtime is required to allocate memory and construct $\mathbf{A}$ or to construct the normal matrix directly.

\begin{figure}[]
    \centering
    \includegraphics[width=0.9\columnwidth]{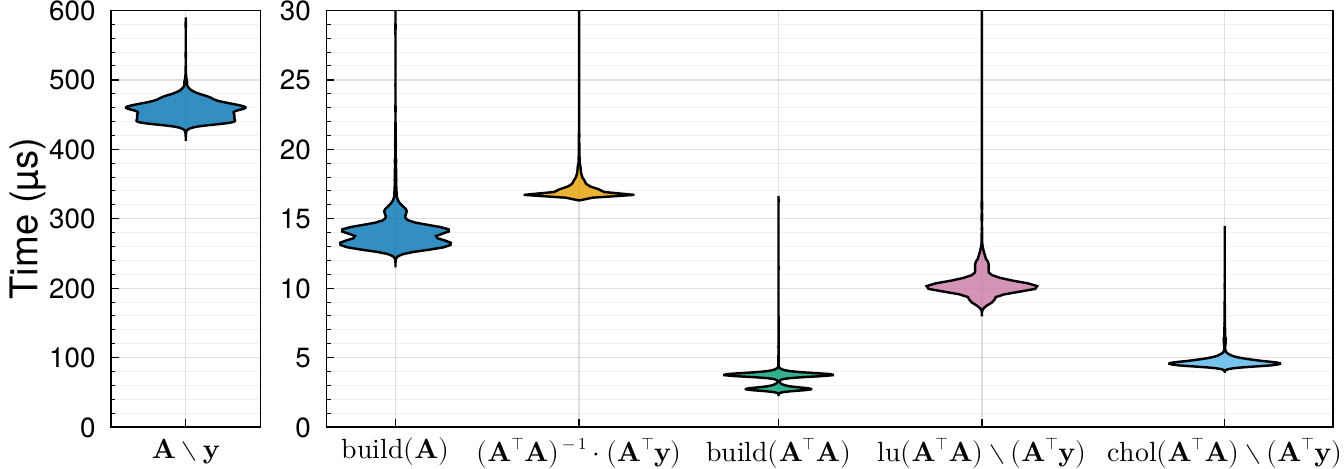}
    \caption{Single core benchmark results on a Ryzen 5950X CPU for two matrix build options and various solver algorithms to calculate the least-square approximation $\mathbf{\hat{x}}$. The $\setminus$ operator performs the linear solution.}
    \label{fig:solver_benchmark}
\end{figure}

In figure~\ref{fig:solver_benchmark}, benchmark results of different methods for solving the equation are shown with the example of figure~\ref{fig:model_function}b as problem definition.
The Julia backslash operator takes about \SI{450}{\micro s}. It serves as a reference for improvement and we can also see that constructing the matrix $\mathbf{A}$ takes about \SI{13}{\micro s}. 
The $\text{build}(\mathbf{A})$ operation is usually limited by memory allocation, especially as the signal length increases.
The runtime compared to the reference can already be drastically reduced to about \SI{17}{\micro s} by solving the equation~\ref{eq:solve_inv} by the inverse instead.
An additional performance improvement can be achieved by calculating a factorization first.
This eliminates the need to compute an inverse matrix, and the backslash operator performs forward and backward substitution to solve the equation.
For this type of problem, where $\mathbf{B = (A^\top A)}$ is the normal matrix and positive definite, the Cholesky factorization is about twice as fast as the LU factorization.
The runtime penalty due to memory allocation can be improved by constructing the normal matrix directly.
The elements of the normal matrix can be derived from the definition of the rectangular matrix in equation~\ref{eq:Atimesx} and the matrix multiplication.
First, we can take advantage of the fact that the normal matrix is symmetric and square, so only the upper or lower half needs to be calculated.
The first element is then
\begin{equation}
    B_{1,1} = N,
\end{equation}
and the first row is 
\begin{equation}
    B_{1,j} = \sum_{k=1}^{K}
    \begin{cases}
        p[k] & \text{if } \; 1 \leq (v[j] - s + k) \leq K \\
        0, & \text{otherwise}  \\
    \end{cases}
\end{equation}
where most row elements are the same as long as a pulse does not extend beyond the signal range.
All other elements of the matrix are equal to the dot product of two pulses shifted relative to each other according to their positions in the matrix, this shift is then $l = |v[j]-v[i]|$.
Again, with the additional condition that pulses may extend beyond the signal range, this gives
\begin{equation}
    B_{i,j} = \sum_{k=1}^{K-l}
    \begin{cases}
        p[k] \cdot p[k+l], & \text{if } \; 1 \leq (v[\text{max}(i,j)] - s + k) \leq K \\
        0, & \text{otherwise}.  \\
    \end{cases}
\end{equation}
As long as the pulse shape does not change, the results of the dot product of specific shifts can be cached, further improving performance when duplicates occur.

\begin{figure}[]
    \centering
    \includegraphics[width=0.9\columnwidth]{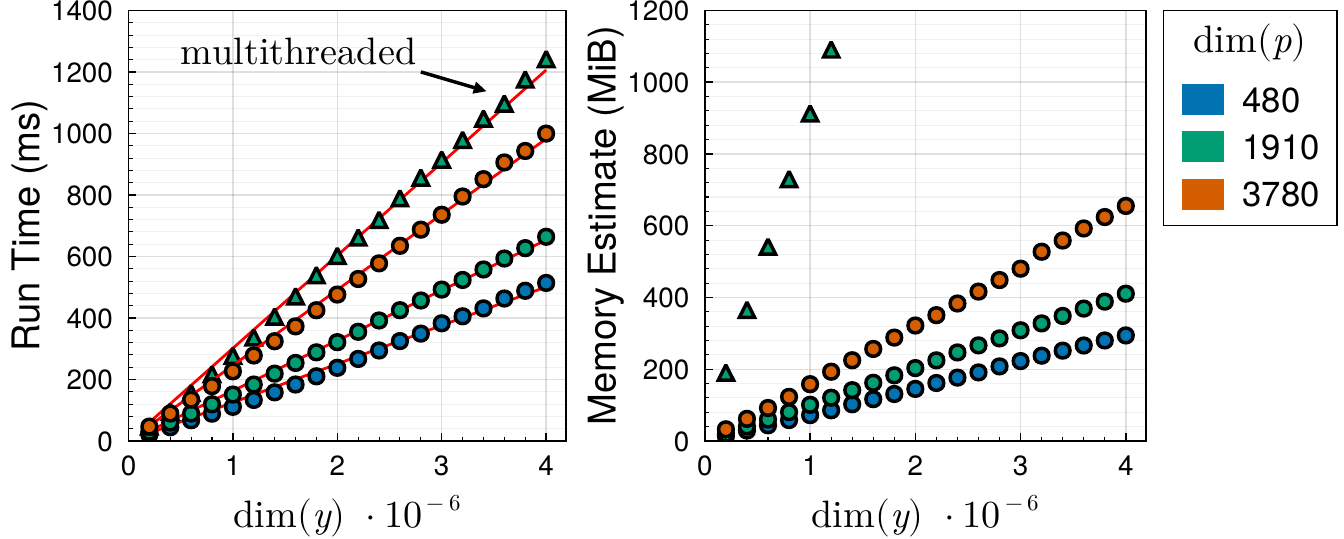}
    \caption{Benchmark results as a function of various signal lengths and three different pulse lengths on a Ryzen 5950X (single core). The pulse count rate of the test signals is \SI{20}{\mega \hertz} at a sampling rate of \SI{5}{GS/s}. The multithreading benchmark (triangles) processed eight signals of the same length in parallel, resulting in a speedup of approximately 4.3.}
    \label{fig:algorithm_benchmark}
\end{figure}

Figure~\ref{fig:algorithm_benchmark} shows the results of a benchmark that tests the entire algorithm as a function of signal length. The runtime is given as the median of 20 samples.
The test signal to be processed for this benchmark is generated at a sampling rate of 5GS/s with white noise superimposed, similar to figure~\ref{fig:model_function}.
The pulse count rate is \SI{20}{MHz} and the pulses were distributed in time according to a Poisson distribution.
The algorithm was set to restart twice using the residual, and the inner pulse optimizer loop was set to run three times per restart (figure~\ref{fig:algorithm_flowchart}).
These settings result in nine least squares solutions being calculated per signal.

The runtime in figure~\ref{fig:algorithm_benchmark} scales almost linearly with increasing signal length, as visualized by fitted straight red lines.
For the longest signal length of \SI{4}{MSamples}, an average of 16k SiPM pulses are processed.
Three different pulse lenghts were simulated with \SI{480}{Samples}, \SI{1910}{Samples} and \SI{3780}{Samples}.
For a given signal length, the run time as a function of the pulse length increases faster than linearly.
The average number of pulses overlapping a given pulse increases, reducing the sparsity of the normal matrix and increasing the computational complexity in a non-linear manner.
A multithreading benchmark \cite{juliadoc} of eight signals of the same length is done for the pulse length of \SI{1910}{Samples}.
While the computational load increased by a factor of eight, the runtime increased by less than a factor of two. 
This gives a speedup of about 4.3 for this configuration and on this machine.

Allocated memory also scales almost linearly as a function of signal length, as shown in Figure~\ref{fig:algorithm_benchmark}b, and appears to correlate with runtime.
This suggests that further optimization of the algorithm in terms of memory allocation may also improve run time.
In the case of multi-threading, the allocated memory increased linearly with the number of additional threads.

\subsection{Simulated results}

The pulse shape of a SiPM may depend on the size of the microcell, the manufacturer, the signal chain including preamplifiers, or other parameters.
A general representation of a pulse can be modeled by an exponentially modified Gaussian distribution:
\begin{equation}
    \label{eq:exp_Gaussian}
    h(x; A, \mu, \sigma, \tau) = A \cdot \frac{\sigma}{\tau} \sqrt{\frac{\pi}{2}} \exp\left( \frac{1}{2} \left(\frac{\sigma}{\tau}\right)^2  - \frac{x-\mu}{\tau}\right) \mathrm{erfc}\left(\frac{1}{\sqrt{2}} \left(\frac{\sigma}{\tau}- \frac{x-\mu}{\sigma}\right) \right),
\end{equation}
where $A$ is the amplitude, $\sigma^2$ is the variance, $\mu$ is the mean, erfc is the complementary error function and $\tau$ is the exponent relaxation time.
However, SiPM pulses often consist of two exponential decays, a fast and a slow recovery \cite{acerbi_understanding_2019}. Two distributions can be superimposed and the new model function is then
\begin{equation}
    \label{eq:pulse_shape_function}
    h(t; A_\mathrm{fast}, A_\mathrm{slow}, \sigma, \tau_\mathrm{fast}, \tau_\mathrm{slow}) = h_\mathrm{fast}(t; A_\mathrm{fast}, 0, \sigma, \tau_\mathrm{fast}) + h_\mathrm{slow}(t; A_\mathrm{slow}, 0, \sigma, \tau_\mathrm{slow}).
\end{equation}

\begin{figure}[]
    \centering
    \includegraphics[width=1\columnwidth]{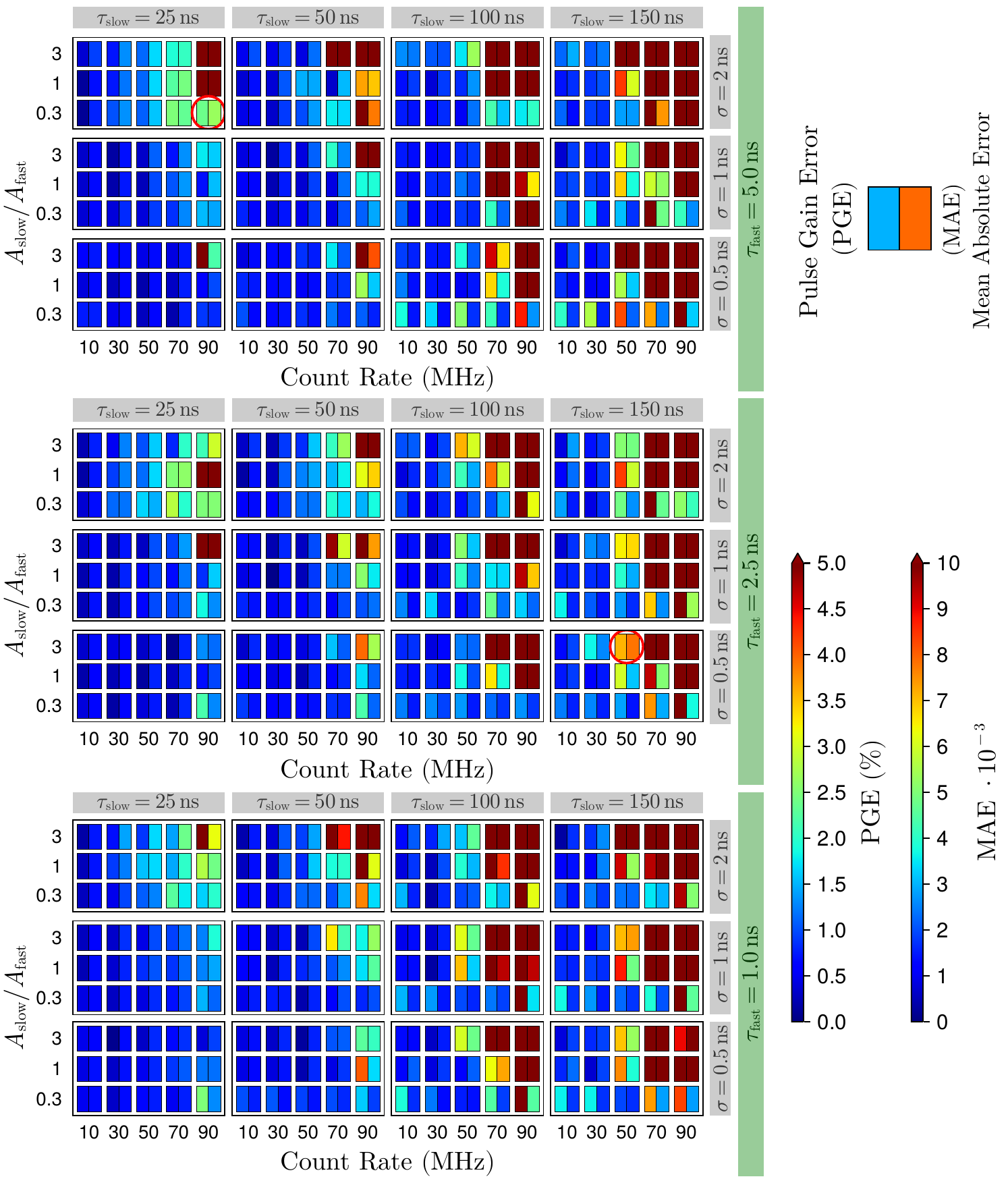}
    \caption{Pulse shape detection results as a function of the input signal count rate for different pulse shape parameters (Eq.~\refeq{eq:pulse_shape_function}). The signal sections and the pulse shape results of the two marked configurations (red circle) are shown in figure~\ref{fig:shape_validation}.}
    \label{fig:pulse_shape_matrix}
\end{figure}

The performance of the pulse shape recognition is tested for different pulse count rates and various combinations of $\sigma$, $\tau_\mathrm{fast}$,  $\tau_\mathrm{slow}$, $A_\mathrm{fast}$ and $A_\mathrm{slow}$.
Two performance figures are used to evaluate the results.
The first figure evaluates the relative difference between the true pulse vector area and the actual area. 
Here called the Pulse Gain Error (PGE) and given as
\begin{equation}
    \mathrm{PGE} = \left| \frac{\mathbf{\sum p_A - \sum p_E }}{\mathbf{\sum p_E}} \right|,
\end{equation}
where $\mathbf{p_E}$ is the expected or true pulse used to generate the test signals and $\mathbf{p_A}$ is the actual pulse that is detected by the pulse shape algorithm.
Whilst the PGE can be low even for very different shapes, the second performance figure should indicate how close the shape of the true pulse is to the shape of the detected pulse.
We use the mean absolute error (MAE) as the second performance figure, defined here as
\begin{equation}
    \mathrm{MAE} = \frac{\sum_{i=1}^{K} \left| \mathbf{p_{A}}[i] - \mathbf{p_{E}}[i] \right| }{K},
\end{equation}
where the pulses are aligned such that $\mathrm{arg\,max}(\mathbf{|p_E|}) = \mathrm{arg\,max}(\mathbf{|p_A|})$ and their pulse heights are normalized to $\mathrm{max}(\mathbf{|p_E|}) = \mathrm{max}(\mathbf{|p_A|}) = 1$.
Good recognition should result in a low value for both errors.

Figure~\ref{fig:pulse_shape_matrix} shows the results of all parameter combinations tested at pulse count rates from \SI{10}{MHz} to \SI{90}{MHz}.
The signals are generated with a sampling rate of \SI{5}{GS/s}, a pulse amplitude of \SI{-7}{mV} and normally distributed white noise with $\sigma = \SI{0.6}{mV}$.
The parameter values should cover the most common SiPM pulse shapes, e.g. the long recovery time of a KETEK PM3350 ($\tau_\mathrm{slow} \approx\SI{130}{ns}$) \cite{PM3350}, a medium pulse length of a Broadcom NUV-MT ($\tau_\mathrm{slow} \approx \SI{55}{ns}$) \cite{NUVMT} or a rather fast response of a SiPM with a small microcell size like a Hamamatsu S14160-1315PS ($\tau_\mathrm{slow} \approx\SI{25}{ns}$) \cite{HPK}.
The same algorithm settings are used for all pulse shapes and count rates, i.e. a threshold of \SI{-2.5}{mV} and a pulse position optimization range of \SI{10}{Samples} before the peak to \SI{15}{Samples} after the peak.
The maximum number of iterations for pulse shape detection is set to one hundred.
The length of the signal is adapted to the count rate so that 2000 pulses can be evaluated for each configuration. 
Five such randomly generated signals are processed per configuration.
The averaged results are shown in figure~\ref{fig:pulse_shape_matrix}.
The color range limits are \SI{5}{\percent} and \SI{1}{\percent} for the PGE and the MAE, respectively.
Results above these thresholds are displayed in the same color as the threshold.

\begin{figure}[]
    \centering
    \includegraphics[width=0.9\columnwidth]{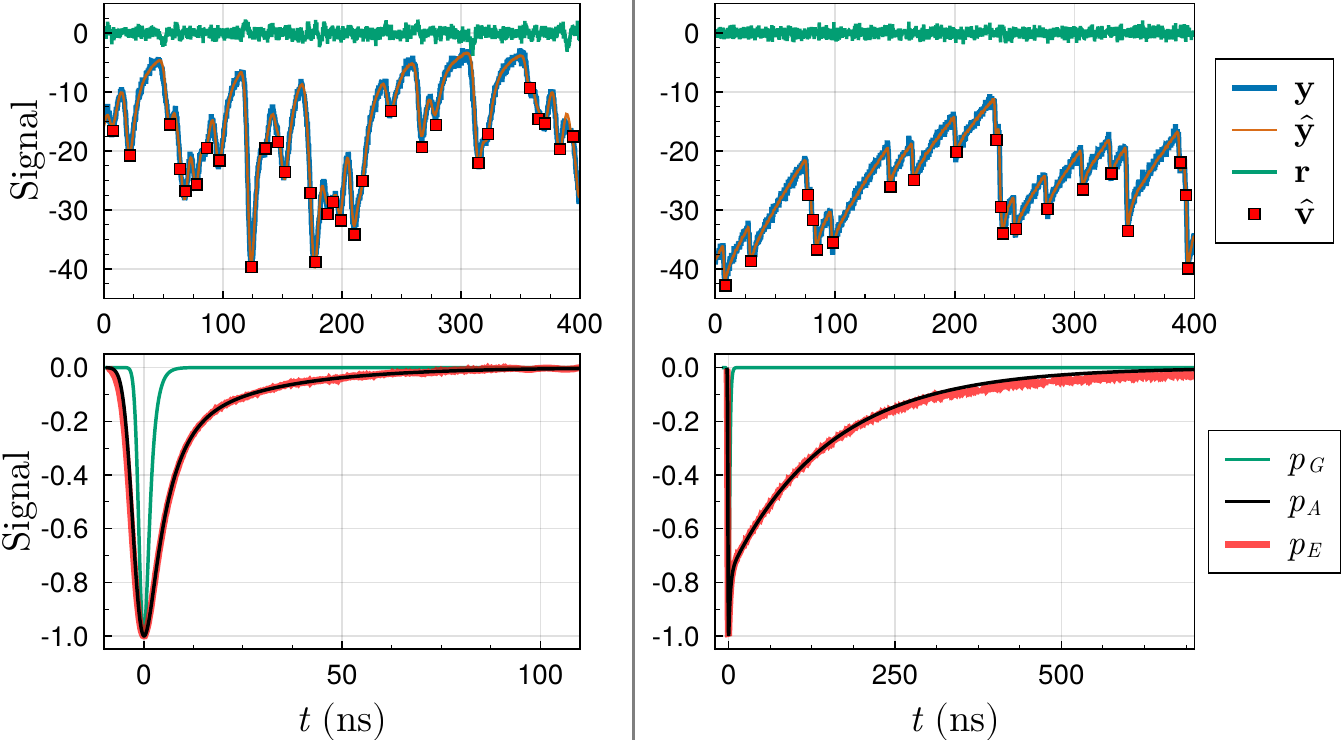}
    \caption{Signals processed by the pulse detection algorithm of two example configurations marked with a red circle in figure~\ref{fig:pulse_shape_matrix}. \textbf{(left)}~$\sigma=\SI{2}{ns}$, $\tau_\mathrm{fast}=\SI{5}{ns}$, $\tau_\mathrm{slow}=\SI{25}{ns}$, $A_\mathrm{slow}/A_\mathrm{fast} =0.3$ and the count rate is \SI{90}{MHz}. \textbf{(right)}~$\sigma=\SI{0.5}{ns}$, $\tau_\mathrm{fast}=\SI{2.5}{ns}$, $\tau_\mathrm{slow}=\SI{150}{ns}$, $A_\mathrm{slow}/A_\mathrm{fast} =3$ and the count rate is \SI{50}{MHz}.
    \textbf{(top)}~Section of the generated and processed signal. \textbf{(bottom)}~Inital guess ($p_G$), true pulse shape ($p_A$) and estimated pulsed shape ($p_E$).}
    \label{fig:shape_validation}
\end{figure}

For the two configurations marked with red circles, a section of a processed signal and its detection results are shown in figure~\ref{fig:shape_validation} to provide context for the results.
While the detection quality of the two selected configurations is in the medium to high error range, the majority of the pulse shapes are still well estimated despite the high count rates.

Overall, the pulse shape detection algorithm works very well for all pulse shapes tested up to \SI{30}{MHz}, with a performance largely independent of the values for $\tau_\mathrm{fast}$.
Increasing the count rate further shows that the failure rate is a function of $\tau_\mathrm{slow}$, and a higher ratio of $A_\mathrm{slow}/A_\mathrm{fast}$ is less favorable.
The increasing overlap of pulses with increasing count rate could explain this observation, since the pulse shape that minimizes the residual may no longer be unique, resulting in a divergence from the true pulse shape.
Remarkably, the presented algorithm can recover some pulse shapes even up to a count rate of \SI{90}{MHz}.

\subsection{Experimental results}

In the previous section, generated signals were used to validate the performance of the presented algorithm, while in this section, signals from a SiPM coupled to a pre-amplifier and digitized with an oscilloscope are used to demonstrate an application.

Ideally, a SiPM is a photon detector with a linear charge response to the number of photons incident on the sensitive area.
However, the actual response deviates from this ideal case due to nuisance effects \cite{klanner_characterisation_2019}, such as dark counts, prompt and delayed optical crosstalk or afterpulses.
Undesirable counts can also be caused by ambient light or radiation damage.
Evaluation is also affected by electronic noise.
These effects can be observed in a charge spectrum, where the x-axis represents the number of Geiger discharges or the number of triggered cells and the y-axis represents the number of counts.
Such a charge spectrum is typically acquired by signal integration with a given gate length.
This integration window can be placed around the expected arrival time of the photons, or a synchronization signal if artificial illumination such as a laser source is used.
An alternative evaluation uses the pulse amplitudes, however, such spectra are typically significantly affected by the electronic noise of the front end, by its bandwidth and by peak oscillations \cite{acerbi_understanding_2019}.

\begin{figure}[]
    \centering
    \includegraphics[width=.9\columnwidth]{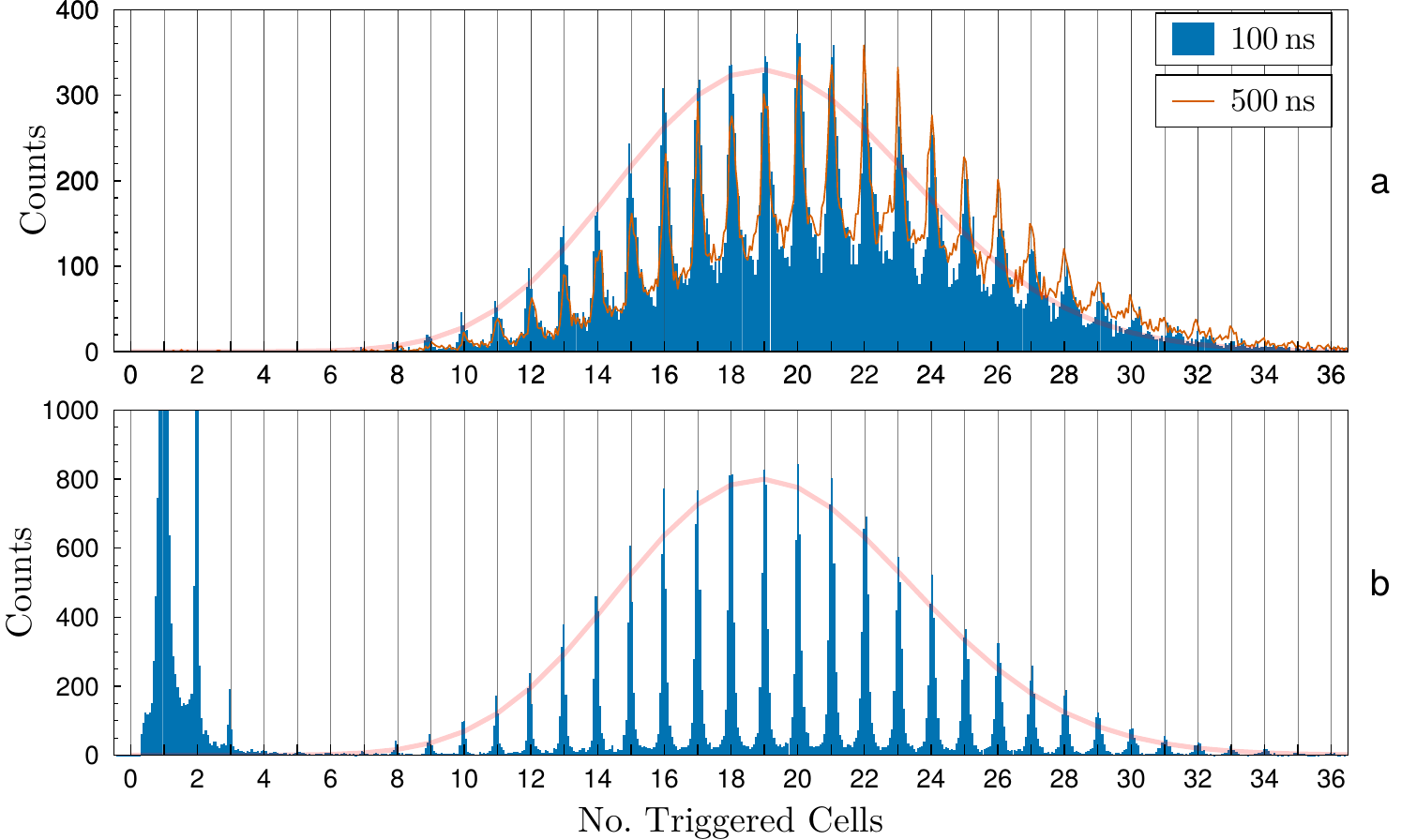}
    \caption{ \textbf{(a)}~Spectra obtained using an integration gate around the laser synchronization signal. Two different integration windows (\SI{100}{ns} and \SI{500}{ns}) are shown. \textbf{(b)}~For comparison, the spectrum determined by the fitting algorithm. The contribution of the laser light and the ambient light is well separated. 
    An upscaled Poisson distribution (red) with $\lambda=19.4$ is plotted for both methods as a visual guide.
    }
    \label{fig:nuvmit_spectra}
\end{figure}

To demonstrate the capabilities of our algorithm, we illuminated a 1x\SI{1}{mm^2} Broadcom NUV-MT \cite{NUVMT} with a laser to obtain charge spectra.
The pulse width of the laser was less than \SI{100}{ps}.
In the measurement environment, ambient light was incident on the SiPM and without laser light, a count rate of approximately \SI{2.5}{MHz} was measured.

Figure \ref{fig:nuvmit_spectra}a shows the charge spectrum obtained by using charge integration. The integration window lengths were \SI{100}{ns} and \SI{500}{ns}.
Each spectrum consists of 40k pulses.
The integration window was aligned using the laser synchronization signal.

In figure~\ref{fig:nuvmit_spectra}b, the same raw data were processed with the algorithm of this paper using the following procedure.
The pulse vector is first estimated using a fraction of the data, in this case, the signal vector contained about 500 pulses. Then the entire data is processed using this estimated pulse vector.
If the detected pulse vector is normalized before processing so that $\mathrm{max}(\mathbf{|p|})=1$, then the pulse integral of each pulse is equal to the integral of the normalized pulse shape times the detected amplitude.
The algorithm processes all pulses, regardless of whether they are generated by laser light, ambient light, or thermal generation (dark counts). 
This approach was used to create the lower spectrum in figure~\ref{fig:nuvmit_spectra}b. 

\begin{figure}[]
    \centering
    \includegraphics[width=0.9\columnwidth]{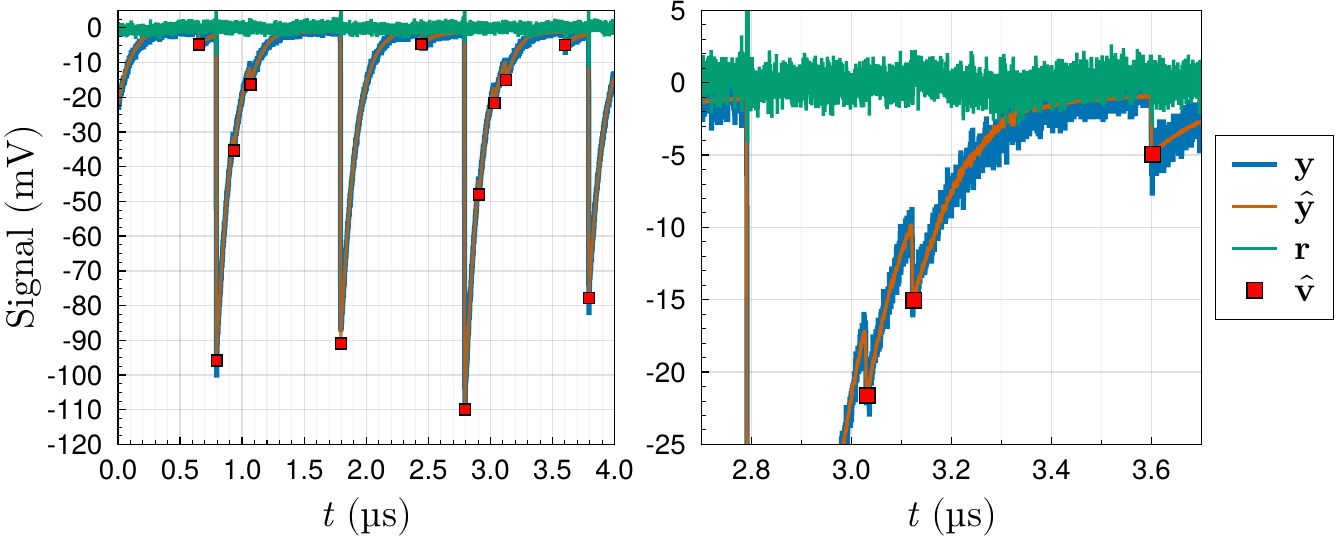}
    \caption{Processed example of the transient signal used to determine the spectra in figure~\ref{fig:nuvmit_spectra}. The count rate due to ambient light is approximately $\SI{2.5}{MHz}$.}
    \label{fig:nuvmt-transient}
\end{figure}

The spectra of figure~\ref{fig:nuvmit_spectra} are quite different. First, since the algorithm evaluates all pulses, counts below five triggered cells appear.
Then, compared to the classical integration, we see that the peaks related to the laser illumination are much narrower.
For example, the peak-to-valley ratio, at 19 triggered cells, is greater than 40 for the algorithm and less than four for the spectrum with the \SI{100}{ns} integration window, improving the peak-to-valley ratio by an order of magnitude.
In addition, peaks up to 34 triggered cells are well resolved in the lower spectrum and the spectrum shows very good linearity.
A third observation can be done by plotting an upscaled Poisson distribution with $\lambda=19.4$ as an overlay on all spectra.
If the number of incoming photons follows a Poisson distribution, then the number of counts in the peaks should follow the same distribution for an ideal photon detector.
It can be seen that the peaks of the algorithm spectrum follow the Poisson distribution very well, while the integrated spectra deviate from this ideal case as the number of triggered cells increases.
This distortion can be caused by correlated noise, such as crosstalk, or by independent noise events falling within the integration window. 
Since our algorithm determines the charge from the amplitude of the pulse, we can eliminate the contribution of noise and delayed correlated pulses compared to the classical integration approach.
As an example, figure~\ref{fig:nuvmt-transient} shows the acquired data along with the calculated pulses and signal from our algorithm.
We can clearly see the peaks from the laser as well as additional trailing edge pulses caused by dark noise events or delayed correlated events. In classical charge integration, the peak-to-valley ratio of the spectrum decreases when such additional events are partially integrated.
In the case of the \SI{500}{ns} integration window, these effects are even more pronounced.

\section{Conclusion}
\label{sec:conlusion}

In this paper, we presented a novel algorithm that uses multiple linear regression to iteratively process transient SiPM signals and an extended version to automatically detect the pulse shape.
Successful pulse shape detection of various SiPM-like signal shapes has been demonstrated for count rates up to \SI{90}{MHz}.
Furthermore, we have shown that the peak-to-valley ratio of a SiPM spectrum can be improved by an order of magnitude using this algorithm.
These potential benefits all come with good performance in terms of runtime and memory requirements, however, it relies on linear algebra libraries.
This requirement is not a limiting factor for implementation on integrated circuits such as field-programmable gate arrays (FPGAs).
It has been shown that linear algebra can be implemented on FPGAs \cite{gonzalez_lapackrc_2009}, and for sparse Cholesky factorization, even performance similar to that on GPUs has been reported \cite{sun_sparse_2017}.
Linear solver libraries for FPGAs are available also commercially \cite{xilinx_vitis}.

This signal processing method applies to various SiPM designs operated under different conditions, an evaluation could benefit especially when high count rates occur due to high temperatures, radiation damage, or ambient light.
Another advantage is the very good suppression of electronic noise or the filtering of SiPM excess counts as demonstrated in the spectrum comparison.
In addition, the algorithm automatically adapts to changes in pulse shape.
For example, this may occur when the temperature changes and a characterization of the gain as a function of temperature can benefit from this feature.
Several parameters of the algorithm can be adjusted to suit specific conditions, however, very good results can be achieved using the same settings to process a wide variety of pulse shapes.
Another area of application may be the improvement of timing measurement.

\bibliographystyle{IEEEtran.bst}
\bibliography{mybibfile}

\begin{thebibliography}{10}
\providecommand{\url}[1]{#1}
\csname url@samestyle\endcsname
\providecommand{\newblock}{\relax}
\providecommand{\bibinfo}[2]{#2}
\providecommand{\BIBentrySTDinterwordspacing}{\spaceskip=0pt\relax}
\providecommand{\BIBentryALTinterwordstretchfactor}{4}
\providecommand{\BIBentryALTinterwordspacing}{\spaceskip=\fontdimen2\font plus
\BIBentryALTinterwordstretchfactor\fontdimen3\font minus
  \fontdimen4\font\relax}
\providecommand{\BIBforeignlanguage}[2]{{%
\expandafter\ifx\csname l@#1\endcsname\relax
\typeout{** WARNING: IEEEtran.bst: No hyphenation pattern has been}%
\typeout{** loaded for the language `#1'. Using the pattern for}%
\typeout{** the default language instead.}%
\else
\language=\csname l@#1\endcsname
\fi
#2}}
\providecommand{\BIBdecl}{\relax}
\BIBdecl

\bibitem{gundacker_silicon_2020}
\BIBentryALTinterwordspacing
S.~Gundacker and A.~Heering, ``The silicon photomultiplier: fundamentals and
  applications of a modern solid-state photon detector,'' \emph{Physics in
  Medicine \& Biology}, vol.~65, no.~17, p. 17TR01, Sep. 2020. [Online].
  Available: \url{https://iopscience.iop.org/article/10.1088/1361-6560/ab7b2d}
\BIBentrySTDinterwordspacing

\bibitem{lecoq_sipm_2021}
\BIBentryALTinterwordspacing
P.~Lecoq and S.~Gundacker, ``\BIBforeignlanguage{en}{{SiPM} applications in
  positron emission tomography: toward ultimate {PET} time-of-flight
  resolution},'' \emph{\BIBforeignlanguage{en}{The European Physical Journal
  Plus}}, vol. 136, no.~3, p. 292, Mar. 2021. [Online]. Available:
  \url{https://link.springer.com/10.1140/epjp/s13360-021-01183-8}
\BIBentrySTDinterwordspacing

\bibitem{santangelo_si_2016}
\BIBentryALTinterwordspacing
M.~F. Santangelo, E.~L. Sciuto, S.~A. Lombardo, A.~C. Busacca, S.~Petralia,
  S.~Conoci, and S.~Libertino, ``\BIBforeignlanguage{en}{Si {Photomultipliers}
  for {Bio}-{Sensing} {Applications}},'' \emph{\BIBforeignlanguage{en}{IEEE
  Journal of Selected Topics in Quantum Electronics}}, vol.~22, no.~3, pp.
  335--341, May 2016. [Online]. Available:
  \url{http://ieeexplore.ieee.org/document/7343739/}
\BIBentrySTDinterwordspacing

\bibitem{garutti_silicon_2011}
\BIBentryALTinterwordspacing
E.~Garutti, ``Silicon photomultipliers for high energy physics detectors,''
  \emph{Journal of Instrumentation}, vol.~6, no.~10, pp. C10\,003--C10\,003,
  Oct. 2011. [Online]. Available:
  \url{https://iopscience.iop.org/article/10.1088/1748-0221/6/10/C10003}
\BIBentrySTDinterwordspacing

\bibitem{bilik_comparative_2022}
\BIBentryALTinterwordspacing
I.~Bilik, ``\BIBforeignlanguage{en}{Comparative {Analysis} of {Radar} and
  {Lidar} {Technologies} for {Automotive} {Applications}},''
  \emph{\BIBforeignlanguage{en}{IEEE Intelligent Transportation Systems
  Magazine}}, pp. 2--27, 2022. [Online]. Available:
  \url{https://ieeexplore.ieee.org/document/9760734/}
\BIBentrySTDinterwordspacing

\bibitem{acerbi_high-density_2018}
\BIBentryALTinterwordspacing
F.~Acerbi, G.~Paternoster, A.~Gola, V.~Regazzoni, N.~Zorzi, and C.~Piemonte,
  ``High-{Density} {Silicon} {Photomultipliers}: {Performance} and {Linearity}
  {Evaluation} for {High} {Efficiency} and {Dynamic}-{Range} {Applications},''
  \emph{IEEE Journal of Quantum Electronics}, vol.~54, no.~2, pp. 1--7, Apr.
  2018. [Online]. Available: \url{http://ieeexplore.ieee.org/document/8281487/}
\BIBentrySTDinterwordspacing

\bibitem{acerbi_understanding_2019}
\BIBentryALTinterwordspacing
F.~Acerbi and S.~Gundacker, ``\BIBforeignlanguage{en}{Understanding and
  simulating {SiPMs}},'' \emph{\BIBforeignlanguage{en}{Nuclear Instruments and
  Methods in Physics Research Section A: Accelerators, Spectrometers, Detectors
  and Associated Equipment}}, vol. 926, pp. 16--35, May 2019. [Online].
  Available:
  \url{https://linkinghub.elsevier.com/retrieve/pii/S0168900218317704}
\BIBentrySTDinterwordspacing

\bibitem{stein_x-ray_1996}
\BIBentryALTinterwordspacing
J.~Stein, F.~Scheuer, W.~Gast, and A.~Georgiev, ``\BIBforeignlanguage{en}{X-ray
  detectors with digitized preamplifiers},''
  \emph{\BIBforeignlanguage{en}{Nuclear Instruments and Methods in Physics
  Research Section B: Beam Interactions with Materials and Atoms}}, vol. 113,
  no. 1-4, pp. 141--145, Jun. 1996. [Online]. Available:
  \url{https://linkinghub.elsevier.com/retrieve/pii/0168583X95014179}
\BIBentrySTDinterwordspacing

\bibitem{nakhostin_signal_2018}
M.~Nakhostin, \emph{Signal processing for radiation detectors}.\hskip 1em plus
  0.5em minus 0.4em\relax Hoboken, NJ, USA: Wiley, 2018.

\bibitem{engelmann_sipm_2018}
E.~Engelmann, ``Sipm noise measurement with waveform analysis,'' Schwetzingen,
  Germany, 2018.

\bibitem{piemonte_development_2012}
\BIBentryALTinterwordspacing
C.~Piemonte, A.~Ferri, A.~Gola, A.~Picciotto, T.~Pro, N.~Serra, A.~Tarolli, and
  N.~Zorzi, ``Development of an automatic procedure for the characterization of
  silicon photomultipliers,'' \emph{2012 IEEE Nuclear Science Symposium and
  Medical Imaging Conference Record (NSS/MIC)}, pp. 428--432, Oct. 2012.
  [Online]. Available: \url{http://ieeexplore.ieee.org/document/6551141/}
\BIBentrySTDinterwordspacing

\bibitem{gola_dled_2012}
\BIBentryALTinterwordspacing
A.~Gola, C.~Piemonte, and A.~Tarolli, ``The {DLED} {Algorithm} for {Timing}
  {Measurements} on {Large} {Area} {SiPMs} {Coupled} to {Scintillators},''
  \emph{IEEE Transactions on Nuclear Science}, vol.~59, no.~2, pp. 358--365,
  Apr. 2012. [Online]. Available:
  \url{http://ieeexplore.ieee.org/document/6175972/}
\BIBentrySTDinterwordspacing

\bibitem{putignano_non-linear_2012}
\BIBentryALTinterwordspacing
M.~Putignano, A.~Intermite, and C.~P. Welsch, ``\BIBforeignlanguage{en}{A
  non-linear algorithm for current signal filtering and peak detection in
  {SiPM}},'' \emph{\BIBforeignlanguage{en}{Journal of Instrumentation}},
  vol.~7, no.~08, pp. P08\,014--P08\,014, Aug. 2012. [Online]. Available:
  \url{https://iopscience.iop.org/article/10.1088/1748-0221/7/08/P08014}
\BIBentrySTDinterwordspacing

\bibitem{bychkova_radiation_2022}
\BIBentryALTinterwordspacing
O.~Bychkova, P.~Parygin, E.~Garutti, A.~Kaminsky, S.~Martens, E.~Popova,
  J.~Schwandt, and A.~Stifutkin, ``\BIBforeignlanguage{en}{Radiation hardness
  study using {SiPMs} with single-cell readout},''
  \emph{\BIBforeignlanguage{en}{Nuclear Instruments and Methods in Physics
  Research Section A: Accelerators, Spectrometers, Detectors and Associated
  Equipment}}, vol. 1031, p. 166533, May 2022. [Online]. Available:
  \url{https://linkinghub.elsevier.com/retrieve/pii/S0168900222001395}
\BIBentrySTDinterwordspacing

\bibitem{brakel2014}
\BIBentryALTinterwordspacing
J.~v. Brakel, ``\BIBforeignlanguage{en}{Robust peak detection algorithm using
  z-scores},'' Stack Overflow, 2014. [Online]. Available:
  \url{https://stackoverflow.com/questions/22583391/peak-signal-detection-in-realtime-timeseries-data/22640362#22640362}
\BIBentrySTDinterwordspacing

\bibitem{strutz_data_2015}
T.~Strutz, \emph{Data fitting and uncertainty: a practical introduction to
  weighted least squares and beyond}.\hskip 1em plus 0.5em minus 0.4em\relax
  New York, NY: Springer Berlin Heidelberg, 2015.

\bibitem{strang_introduction_2016}
G.~Strang, \emph{\BIBforeignlanguage{eng}{Introduction to linear algebra}},
  5th~ed.\hskip 1em plus 0.5em minus 0.4em\relax Wellesley: Cambridge press,
  2016.

\bibitem{bezanson_julia_2017}
\BIBentryALTinterwordspacing
J.~Bezanson, A.~Edelman, S.~Karpinski, and V.~B. Shah, ``Julia: A fresh
  approach to numerical computing,'' \emph{SIAM review}, vol.~59, no.~1, pp.
  65--98, 2017. [Online]. Available: \url{https://doi.org/10.1137/141000671}
\BIBentrySTDinterwordspacing

\bibitem{juliadoc}
``Julia {D}ocumentation,'' https://docs.julialang.org/en/v1/.

\bibitem{PM3350}
\BIBentryALTinterwordspacing
``Ketek - silicon photomultiplier pm3350,'' KETEK. [Online]. Available:
  \url{https://www.ketek.net/wp-content/uploads/2017/01/KETEK-PM3325-EB-PM3350-EB-Datasheet.pdf}
\BIBentrySTDinterwordspacing

\bibitem{NUVMT}
\BIBentryALTinterwordspacing
``Broadcom - 2×1 nuv-mt silicon photomultiplier array,'' Broadcom. [Online].
  Available:
  \url{https://www.broadcom.com/products/optical-sensors/silicon-photomultiplier-sipm/afbr-s4n66p024m}
\BIBentrySTDinterwordspacing

\bibitem{HPK}
\BIBentryALTinterwordspacing
``Hamamatsu - mppc s14160 series,'' Hamamatsu. [Online]. Available:
  \url{https://www.hamamatsu.com/eu/en/product/optical-sensors/mppc/mppc_mppc-array.html}
\BIBentrySTDinterwordspacing

\bibitem{klanner_characterisation_2019}
\BIBentryALTinterwordspacing
R.~Klanner, ``\BIBforeignlanguage{en}{Characterisation of {SiPMs}},''
  \emph{\BIBforeignlanguage{en}{Nuclear Instruments and Methods in Physics
  Research Section A: Accelerators, Spectrometers, Detectors and Associated
  Equipment}}, vol. 926, pp. 36--56, May 2019. [Online]. Available:
  \url{https://linkinghub.elsevier.com/retrieve/pii/S0168900218317091}
\BIBentrySTDinterwordspacing

\bibitem{gonzalez_lapackrc_2009}
\BIBentryALTinterwordspacing
J.~Gonzalez and R.~C. Núñez, ``{LAPACKrc}: Fast linear algebra
  kernels/solvers for {FPGA} accelerators,'' vol. 180, p. 012042. [Online].
  Available:
  \url{https://iopscience.iop.org/article/10.1088/1742-6596/180/1/012042}
\BIBentrySTDinterwordspacing

\bibitem{sun_sparse_2017}
\BIBentryALTinterwordspacing
Y.~Sun, H.~Liu, and T.~Zhou, ``Sparse cholesky factorization on {FPGA} using
  parameterized model,'' vol. 2017, pp. 1--11. [Online]. Available:
  \url{https://www.hindawi.com/journals/mpe/2017/3021591/}
\BIBentrySTDinterwordspacing

\bibitem{xilinx_vitis}
\BIBentryALTinterwordspacing
``Vitis solver library,'' AMD Xilinx. [Online]. Available:
  \url{https://www.xilinx.com/products/design-tools/vitis/vitis-libraries/vitis-solver.html}
\BIBentrySTDinterwordspacing

\end{thebibliography}

\end{document}